\documentclass[12pt]{article} 
\usepackage[xdvi]{graphicx} 
\usepackage{amssymb}
\usepackage{amsmath} 
\usepackage{epsfig}
\usepackage{psfrag}
 
\begin{document}   
\newcommand{\todo}[1]{{\em \small {#1}}\marginpar{$\Longleftarrow$}}   
\newcommand{\labell}[1]{\label{#1}\qquad_{#1}} 

\vskip 1cm

\begin{flushright}
RH-16-2005
\end{flushright}

\vskip 2cm
\bigskip
\bigskip
\begin{center}
   {\Large \bf A New Perspective on the Nonextremal }

\bigskip

{\Large \bf Enhan\c con Solution }
   \end{center}

\bigskip
\bigskip
\vskip 0.5cm

\centerline{\bf Jessica K. Barrett\footnote{jessica@raunvis.hi.is}}

\bigskip

\centerline{\it Mathematics Division} \centerline{\it Science Institute} \centerline{\it University of
  Iceland} \centerline{\it Dunhaga 3, IS-107 Reykjavik} \centerline{\it Iceland}


\bigskip
\bigskip
\bigskip


\begin{abstract}
We discuss the nonextremal generalisation of the enhan\c con mechanism. We find that the nonextremal shell branch solution does not violate the Weak Energy Condition when the nonextremality parameter is small, in contrast to earlier discussions of this subject. We show that this physical shell branch solution fills the mass gap between the extremal enhan\c con solution and the nonextremal horizon branch solution.
\end{abstract}

\newpage \baselineskip=18pt \setcounter{footnote}{0}

\section{Introduction}     

The enhan\c con was originally proposed in ref. \cite{Johnson:1999qt} as a new way of resolving singularities in string theory. The nonextremal version of the enhan\c con has been studied in refs. \cite{Johnson:2001wm}, \cite{Dimitriadis:2002xd}, \cite{Dimitriadis:2003ya} and \cite{Dimitriadis:2003ur}, and more recently in ref. \cite{Page:2004sh}, motivated by studies of finite temperature systems within AdS/CFT (see also ref. \cite{Bertolini:2002de} for a discussion of nonextremal fractional branes, which are related to the enhan\c con by T-duality). In particular, in ref. \cite{Johnson:2001wm} it was shown that there are two branches of nonextremal enhan\c con solutions consistent with the supergravity equations of motion; one is referred to as the shell branch because it exhibits a shell similar to that of the enhan\c con, and the other as the horizon branch because it has no shell, but a regular horizon. It was suggested in ref. \cite{Dimitriadis:2003ya} that the shell branch solution violates the Weak Energy Condition (WEC), and therefore only the horizon branch is physically viable. However, there is a mass gap between the extremal enhan\c con solution and the extremal limit of the horizon branch solution, and so a puzzle is presented - what is the form of the nonextremal enhan\c con whose mass lies within this mass gap?

We discuss here the violation of the WEC for the nonextremal enhan\c con solution. Our results are in contrast with those of ref. \cite{Dimitriadis:2003ya}. We find that, although the WEC is violated for large values of the nonextremality parameter $r_0$, when $r_0$ is small enough the WEC is not violated.

The outline of this paper is as follows. In section \ref{sec:review} we review the enhan\c con mechanism and its nonextremal generalisation. In section \ref{sec:WEC} we show that the WEC is not violated by the nonextremal shell branch enhan\c con solution for small values of the nonextremality parameter. In section \ref{sec:ADM} we discuss the implications of this result for the mass gap puzzle.

\section{Review of the Enhan\c con Mechanism}
\label{sec:review}

In this section we will review the enhan\c con mechanism of ref. \cite{Johnson:1999qt}, and the nonextremal generalisation of the enhan\c con mechanism, which has been discussed in refs. \cite{Johnson:2001wm}, \cite{Dimitriadis:2002xd}, \cite{Dimitriadis:2003ya} and \cite{Dimitriadis:2003ur}.

The enhan\c con mechanism describes the effect of wrapping $N$ D$p$-branes on a K3 manifold ($p \geq 5$). This induces $N$ negatively charged D$(p-4)$-branes. To be concrete in what follows we will take $p=6$. Then the Einstein frame supergravity solution for this object  is given by
\begin{eqnarray}
g_s^{1/2} ds^2 & = & Z_2^{-5/8}Z_6^{-1/8} \left( -dt^2 + dx_1^2 + dx_2^2 \right) 
+ Z_2^{3/8}Z_6^{7/8} \left( dr^2 + r^2 d\Omega_2^2 \right) \nonumber \\
&& + V^{1/2} Z_2^{3/8} Z_6^{-1/8} ds_{K3}^2 \,\, , \nonumber \\
e^{2\phi} & = & g_s^2 Z_2^{1/2} Z_6^{-3/2} \,\, , \nonumber \\
C_{(3)} & = & (g_sZ_2)^{-1} dt \wedge dx^1 \wedge dx^2 \,\, , \nonumber \\
C_{(7)} & = & (g_sZ_6)^{-1} dt \wedge dx^1 \wedge dx^2 \wedge V_{K3} \,\, ,
\end{eqnarray}
where $r$ is the radial parameter of the directions transverse to all branes, and $ds_{K3}^2$ and $V_{K3}$ are the K3 line element and volume form respectively. The constant $V$ is the volume of the K3 manifold in the limit $r \to \infty$. The harmonic functions $Z_2$ and $Z_6$ are given by
\begin{equation}
Z_2(r) =  1 + \frac{r_2}{r} \,\, , \quad
Z_6(r) =  1 + \frac{r_6}{r} \,\, ,
\end{equation}
where
\begin{equation}
r_2 = - \frac{(2\pi)^4g_sN\alpha'^{5/2}}{2V} \,\, , \quad r_6 = \frac{g_sN\alpha'^{1/2}}{2} \,\, .
\label{eq:r2r6}
\end{equation}
This solution contains a naked singularity at $r_r = -r_2$ called the repulson singularity. This singularity is resolved in string theory by the enhan\c con mechanism, which says that the supergravity solution inside a certain radius $r_e$, the enhan\c con radius, should be replaced by flat space (see ref. \cite{Johnson:2001wm}). The enhan\c con radius is given by
\begin{equation}
r_e = \frac{2 V_*}{V - V_*} r_6 \,\, ,
\end{equation}
where $V_*= (2\pi\sqrt{\alpha'})^4$ is the volume of the K3 manifold at the enhan\c con radius. Note from equation \eqref{eq:r2r6} that we have a relation between $r_2$ and $r_6$,
\begin{equation}
r_2 = -\frac{V_*}{V} r_6 \,\, .
\label{eq:r2}
\end{equation}
Using \eqref{eq:r2} we find that $r_e>r_r$, and so the enhan\c con mechanism has removed the repulson singularity from the solution.

The Einstein frame metric for the nonextremal version of the enhan\c con solution, i.e. for $N$ non-extremal 6-branes wrapped on a K3 manifold with $N$ induced, negatively charged, non-extremal 2-branes, is given by
\begin{eqnarray}
g_s^{1/2} ds^2 & = & Z_2^{-5/8}Z_6^{-1/8} \left( -Kdt^2 + dx_1^2 + dx_2^2 \right) 
+ Z_2^{3/8}Z_6^{7/8} \left( K^{-1} dr^2 + r^2 d\Omega_2^2 \right) \nonumber \\
&& + V^{1/2} Z_2^{3/8} Z_6^{-1/8} ds_{K3}^2 \,\, . \label{eq:nonextremal_metric}
\end{eqnarray}
Again, this metric is only valid outside the enhan\c con radius. The exterior expressions for the dilaton and R-R fields are given by
\begin{eqnarray}
e^{2\phi} & = & g_s^2 Z_2^{1/2} Z_6^{-3/2} \,\, , \nonumber \\
C_{(3)} & = & (g_s\alpha_2Z_2)^{-1} dt \wedge dx^1 \wedge dx^2 \,\, , \nonumber \\
C_{(7)} & = & (g_s\alpha_6Z_6)^{-1} dt \wedge dx^1 \wedge dx^2 \wedge V\epsilon_{K3} \,\, .
\end{eqnarray}
The harmonic functions $Z_2$, $Z_6$ and $K$ are now given by
\begin{equation}
Z_2(r)  =  1 + \frac{\hat{r}_2}{r} \,\, , \quad
Z_6(r)  =  1 + \frac{\hat{r}_6}{r} \,\, , \quad 
K(r)  =  1 - \frac{r_0}{r} \label{eq:ZK} \,\, .
\end{equation}
where
\begin{equation}
\hat{r}_6  =  -\frac{r_0}{2} + \sqrt{r_6^2 + \left(\frac{r_0}{2}\right)^2} \,\, ,  \quad
\alpha_6 = \frac{\hat{r}_6}{r_6} \,\, ,
\label{eq:r6hat}
\end{equation}
and $r_0$ is the nonextremality parameter of the branes (the extremal limit is $r_0 \to 0$).
There are two choices for $\hat{r}_2$ consistent with the equations of motion
\begin{equation}
\hat{r}_2 = -\frac{r_0}{2} \pm \sqrt{r_2^2 + \left(\frac{r_0}{2}\right)^2} \,\, ,  \quad
\alpha_2 = \frac{\hat{r}_2}{r_2} \,\, .
\label{eq:r2hat}
\end{equation}
$r_6$ and $r_2$ in \eqref{eq:r6hat} and \eqref{eq:r2hat} are still given by \eqref{eq:r2r6}. The choice of a plus sign in $\hat{r}_2$ implies that $\hat{r}_2>0$. Then there is no repulson singularity, and therefore no enhan\c con shell. This solution has a horizon at $r = r_0$, and is therefore known as the horizon branch. On the other hand, the choice of a minus sign in $\hat{r}_2$ implies $\hat{r}_2<0$, and therefore the corresponding solution has a repulson singularity, which is corrected by an enhan\c con shell at $r = r_e$, where $r_e$ is given by (see ref. \cite{Johnson:2001wm})
\begin{equation}
r_e = \frac{V_*\hat{r}_6 - V \hat{r}_2}{V - V_*} \,\, .
\label{eq:re}
\end{equation}
This solution is therefore known as the shell branch. 

The nonextremal solution should tend to the extremal solution in the limit $r_0 \to 0$. In this limit the shell branch solution tends to the extremal solution, whereas the horizon branch solution does not. This suggests that the shell branch solution is the correct solution for small values of $r_0$.  However, it was claimed in ref. \cite{Dimitriadis:2003ya} that the nonextremal shell branch solution violates the WEC. We will now review that calculation.

Matching the nonextremal shell branch solution onto a flat geometry at the incision radius $r_i$, and applying the Israel junction conditions, results in the following expression for the stress energy tensor (ref. \cite{Johnson:2001wm})
\begin{equation}
2 \kappa^2 S_{tt} = \frac{1}{\sqrt{G_{rr}}} \left[ \frac{Z'_2}{Z_2} + \frac{Z'_6}{Z_6} + \frac{4}{r_i} \left(1 - \sqrt{\frac{1}{K(r_i)}} \right) \right] G_{tt} \,\, .
\end{equation}
The energy density of the shell is therefore given by
\begin{equation}
\rho \sim - \frac{Z'_2}{Z_2} - \frac{Z'_6}{Z_6} + \frac{4}{r_i} \left(\sqrt{\frac{1}{K(r_i)}} - 1 \right) \,\, .
\end{equation}
For a shell at the enhan\c con radius we can use
\begin{equation}
\frac{Z_2(r_e)}{Z_6(r_e)} = \frac{V_*}{V}
\end{equation}
to write the energy density as follows
\begin{equation}
\rho \sim \frac{1}{r_e} \frac{1}{Z_2(r_e)} \left(\hat{r}_2 + \frac{V_*}{V} \hat{r}_6 \right)
+ \frac{4}{r_e} \left(\sqrt{\frac{1}{K(r_e)}} - 1 \right) \,\, .
\label{eq:rho}
\end{equation}
For a solution to be physical we require that $\rho\geq 0$, i.e. that the solution does not violate the WEC. In ref. \cite{Dimitriadis:2003ya} it is claimed that the shell branch solution has $\rho < 0$, and is therefore unphysical, whenever the supergravity solution is valid, i.e. whenever $V \gg V_*$. We will discuss our results regarding this issue in the next section.

\section{Violation of the Weak Energy Condition}
\label{sec:WEC}

\begin{figure}
\centering
\psfrag{r0}{$r_0$}
\psfrag{rho}{$\rho$}
\epsfig{figure=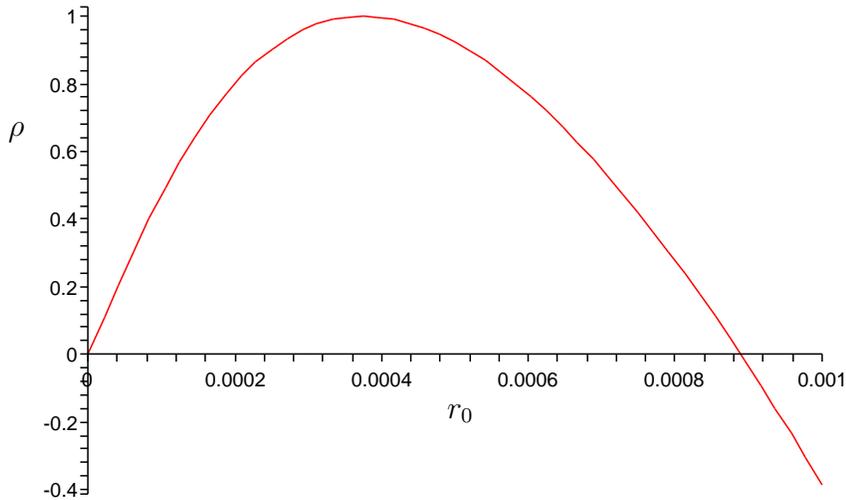,width=.8\textwidth}
\caption{Plot of the expression for the energy density $\rho$ given in equation \eqref{eq:rho} against $r_0$ with $r_6=1$ and $V_*/V=0.0001$. Note that $\rho>0$ for small values of $r_0$.}
\label{fig:energy_density}
\end{figure}

We will show in this section that the WEC is not violated for small enough values of the nonextremality parameter $r_0$. To motivate this calculation we have plotted in figure \ref{fig:energy_density} the expression for $\rho$ from equation \eqref{eq:rho} for small values of $r_0$, taking $r_6=1$ and $V_*/V=0.0001$. At $r_0=0$ we find $\rho=0$, as expected, and when $r_0$ is small we find $\rho>0$.

As was noted in ref. \cite{Dimitriadis:2003ya}, the supergravity solution is only relevant when $V \gg V_*$. We will therefore proceed by expanding all expressions in powers of $V_*/V$. We take $r_0$ to be of the form
\begin{equation}
r_0 = a \frac{V_*}{V} + \cdots \,\, ,
\end{equation} 
where $a$ is independent of $V_*/V$ and $a>0$. We will assume that the parameter $r_6$, given in equation \eqref{eq:r2r6}, is of order unity in the $V_*/V$ expansion, and we use equation \eqref{eq:r2} to express $r_2$ in terms of $r_6$. Then expanding the expressions \eqref{eq:r2hat} (with the minus sign) for $\hat{r}_2$ and \eqref{eq:r6hat} for $\hat{r}_6$ yields
\begin{equation}
\hat{r}_2 = k \frac{V_*}{V} + \cdots \,\, , \quad
\hat{r}_6 = r_6 - \frac{a}{2} \frac{V_*}{V} + \cdots \label{eq:rhat_expanded} \,\, ,
\end{equation}
where
\begin{equation}
k = - \frac{a}{2} - r_6 \sqrt{1 + \frac{a^2}{4r_6^2}} \,\, .
\label{eq:k}
\end{equation}
Using these expressions in the formulae \eqref{eq:re} for $r_e$ and \eqref{eq:ZK} for $Z_2$ and $K$ we find
\begin{equation}
r_e = (r_6-k) \frac{V_*}{V} + \cdots \,\, , \quad 
Z_2(r_e) = \frac{r_6}{r_6-k} + \cdots \,\, , \quad
K(r_e) = 1 - \frac{a}{r_6-k} + \cdots \label{eq:harmonic_expanded}
\end{equation}
Substituting the equations \eqref{eq:rhat_expanded} and \eqref{eq:harmonic_expanded} into the expression for the energy density $\rho$ \eqref{eq:rho}, we find
\begin{equation}
\rho \sim \left(\frac{V_*}{V}\right)^{-1} \frac{1}{r_6-k} \left( \frac{k}{r_6} - 3 + 4 \left(1 - \frac{a}{r_6-k}\right)^{-1/2}\right) + \cdots \,\, .
\label{eq:rho_expanded}
\end{equation}

\begin{figure}
\centering
\psfrag{b}{$b$}
\psfrag{f(b)}{$f(b)$}
\epsfig{figure=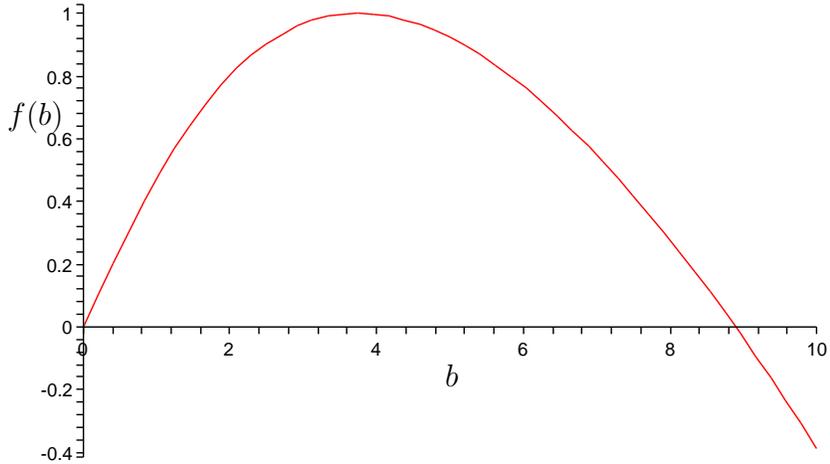,width=.8\textwidth}
\caption{Plot of the leading order term of the energy density $\rho$ in the $V_*/V$ expansion, with $r_6=1$ and $V_*/V=0.0001$.}
\label{fig:f}
\end{figure}

It remains to show that this expression for $\rho$ is positive for some range of values of the parameter $a$. The overall factors in \eqref{eq:rho_expanded} are positive since $k<0$, so we ignore these in what follows. Then setting $a = br_6$, and pulling a factor of $r_6$ out of $\rho$, we can write to leading order
\begin{equation}
\rho \sim \left(-3 -\frac{b}{2} - \sqrt{1 + \frac{b^2}{4}} + 4 \left(1 - b \left(1 + \frac{b}{2} + \sqrt{1 + \frac{b^2}{4}} \right)^{-1} \right)^{-1/2} \right)
\equiv f(b) \,\, ,
\label{eq:rho_b}
\end{equation}
where we have used the definition \eqref{eq:k} of $k$. We now have a leading order expression for $\rho$ as a function only of $b$, which we have defined to be $f(b)$. Plotting $f$ as a function of $b$ we obtain the graph in figure \ref{fig:f} (comparing to figure \ref{fig:energy_density} we find that the two graphs have the same qualitative behaviour, as we would expect). 
We see that $f$, and therefore $\rho$ is positive for $b<\tilde{b}$. In order to find $\tilde{b}$ we seek to solve $f(\tilde{b})=0$. By manipulating this equation, squaring twice to remove the square roots, we find (thanks to some miraculous cancellations of terms) that solutions for $\tilde{b}$ must obey
\begin{equation}
\tilde{b}^2 (-9\tilde{b}+80) = 0 \,\, .
\end{equation}
Substituting the solution to this equation, $\tilde{b} = 80/9$ back into \eqref{eq:rho_b} we find that this is indeed a solution to $f(\tilde{b})=0$.

To summarise, we have shown that the WEC for the shell branch nonextremal enhan\c con solution is not violated for $r_0 \sim a V_*/V$ when $a<80 r_6/9$.

\section{The ADM Mass of the Nonextremal Enhan\c con}
\label{sec:ADM}

The ADM mass of the nonextremal enhan\c con solution described in section \ref{sec:review} is given by (see refs. \cite{Dimitriadis:2002xd} and \cite{Dimitriadis:2003ya})
\begin{equation}
E = \frac{1}{4G} (2r_0 + \hat{r}_2 + \hat{r}_6) \,\, .
\end{equation}
At $r_0 = 0$ the ADM mass of the horizon branch is
\begin{equation}
E_{hb}(r_0=0) = \frac{1}{4G} (r_6+r_2) = \frac{1}{4G} r_6 \left( 1 + \frac{V_*}{V}\right) \,\, ,
\end{equation}
whereas for the extremal enhan\c con solution (i.e. the shell branch solution at $r_0 = 0$) we have
\begin{equation}
E_{ex} = E_{sb} (r_0=0) = \frac{1}{4G}(r_6-r_2) \,\, .
\end{equation}
So there is a mass gap between the mass of the extremal solution and the lowest possible mass of the horizon branch solution. This mass gap has presented a puzzle in the literature (see refs. \cite{Dimitriadis:2003ya} and \cite{Dimitriadis:2003ur}) because the shell branch was thought to be unphysical, and it was unknown what form a solution whose mass lies within the mass gap should take.

We have shown in the previous section that the shell branch is in fact physical for some small values of $r_0$. The question then arises whether or not the mass gap is filled by these physical shell branch solutions. To answer this question, we again expand everything as series in $V_*/V$, and we take
\begin{equation}
r_0 = a \frac{V_*}{V} + \cdots \,\, .
\end{equation}
Then for the ADM mass of the nonextremal enhan\c con solution we find
\begin{equation}
E_{sb}(a) = \frac{1}{4G} \left( r_6 + \left(\frac{3a}{2} + k\right)\frac{V_*}{V} + \cdots \right) \,\, ,
\end{equation}
where $k$ is given by equation \eqref{eq:k}. Again taking $a=br_6$ we find
\begin{equation}
E_{sb}(b) = \frac{r_6}{4G} \left(1 + \left(b - \sqrt{1+\frac{b^2}{4}} \right) \frac{V_*}{V} + \cdots \right) \,\, .
\end{equation}
In the previous section we found that the largest values of $b$ at which the shell branch remained physical was $\tilde{b}=80/9$. At this value of $b$ the shell branch mass is
\begin{equation}
E_{sb}(\tilde{b}) = \frac{r_6}{4G} \left(1 + \frac{39}{9} \frac{V_*}{V} + \cdots \right) > E_{hb}(r_0=0) \,\, .
\end{equation}
We conclude that the mass gap is indeed covered providing the expansion in $V_*/V$ is valid.

\begin{figure}
\centering
\psfrag{r0}{$r_0$}
\psfrag{E}{$E_{\Delta}$}
\epsfig{figure=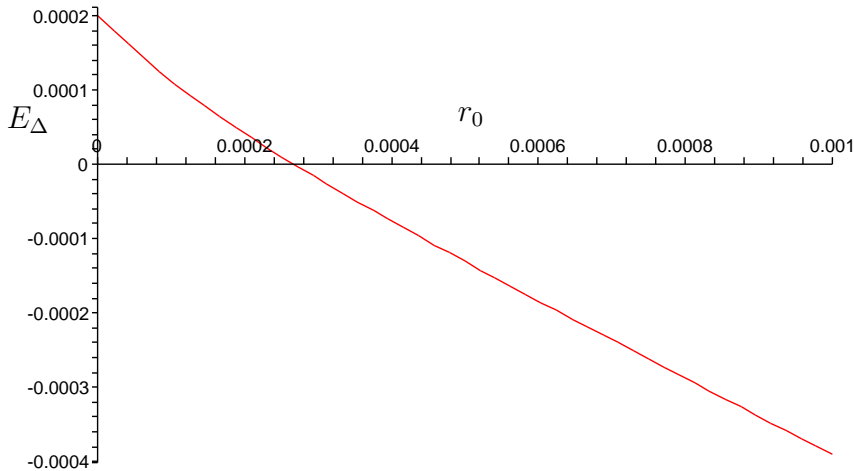,width=.8\textwidth}
\caption{Plot of the energy gap $E_{\Delta}$ between the mass of the shell branch solution with nonextremality parameter $r_0$ and the lowest possible mass of the horizon branch solution, with $r_6=1$ and $V_*/V=0.0001$.}
\label{fig:energy_gap}
\end{figure}

We can also check numerically that the mass gap is covered when we use the exact expressions for $E_{hb}$ and $E_{sb}$. The mass gap between the shell branch solution at $r_0$ and the lowest possible mass of the horizon branch solution is
\begin{equation}
E_{\Delta}(r_0) = E_{hb}(r_0=0) - E_{sb}(r_0) \,\, .
\end{equation}
The graph in figure \ref{fig:energy_gap} shows a plot of $E_{\Delta}$ against $r_0$ for $r_6 = 1$ and $V_*/V = 0.0001$. 
Comparing this graph with that of figure \ref{fig:energy_density} we can see that the mass gap is covered (i.e. $E_{\Delta} < 0 $) for $r_0$ well within the range for which the shell branch solution physical. We have checked this result numerically for various values of $r_6$ and $V_*/V$ (including some for which the expansion in $V_*/V$ is not valid, e.g. $V_*/V=0.1$), and we have found that the mass gap is covered by the physical shell branch solution in all cases.

\section{Conclusions}

We have shown that the nonextremal shell branch enhan\c con solution is physical when the nonextremality parameter $r_0$ is small enough. We have also shown that a supergravity solution of the form given in \eqref{eq:nonextremal_metric} - \eqref{eq:r2hat} exists for all masses above the ADM mass of the extremal enhan\c con. The nonextremal enhan\c con whose mass is close to that of the extremal enhan\c con should take the form of the shell branch solution, because no horizon branch solution exists. This was to be expected because the shell branch solution tends to the extremal solution in the extremality limit $r_0 \to 0$. However, the solution for a nonextremal enhan\c con with a large ADM mass takes the form of the horizon branch solution, because the shell branch solution is unphysical in this region of the parameter space. This was also to be expected because the object with large mass should behave like a black hole, and should therefore have a horizon, as was discussed in ref. \cite{Dimitriadis:2003ya}.

Having identified the form of the nonextremal enhan\c con solution it would be interesting to return to the questions of stability that were addressed in refs. \cite{Dimitriadis:2002xd} and \cite{Dimitriadis:2003ya}. Since the shell branch solution is valid for smaller masses, and the horizon branch for larger masses, we expect a transition from the horizon branch to the shell branch at some value of $r_0$. The stability of the horizon branch solution was tested in refs. \cite{Dimitriadis:2002xd} and \cite{Dimitriadis:2003ya}, but no instabilities were found. However the results we have described here may shed some light on this problem. We now have a clearer idea of the value of $r_0$ at which we expect the transition to occur since the shell branch is unphysical for $r>\tilde{r}_0$, where $\tilde{r}_0 \sim 9 r_6 V_*/V$.

\vskip.5in

\centerline{\bf Acknowledgements}
\medskip    
We thank Clifford Johnson and Larus Thorlacius for useful discussions. This work was supported by a grant from the Icelandic Research Fund.

   
    
\end{document}